\documentclass[a4paper,showpacs,floatfix,twocolumn,pra,superscriptaddress]{revtex4}

\usepackage{graphicx}
\usepackage{amsmath}
\usepackage{color}
\usepackage[english]{babel}
\usepackage{txfonts}
\usepackage{epsfig}
\usepackage{psfrag}

\newcommand{\fref}[1]{Fig.~\ref{#1}}
\newcommand{\eref}[1]{Eq.~(\ref{#1})}
\newcommand{\sref}[1]{Section \ref{#1}}
\newcommand{\aref}[1]{Appendix \ref{#1}}

\newcommand{\sub}[2]{{#1}_{\mbox{\!\! \scriptsize #2}}}

\def\CR{\nonumber\\[0.15cm]}
\newcommand{\etal}{\emph{et al.}}
\newcommand{\kbar}{\hbar_{\rm eff}}

\begin{document}
\title{Dynamical tunnelling with ultracold atoms in magnetic microtraps}

\author{Martin Lenz}
\email{martin.lenz@u-psud.fr}
\affiliation{The University of Queensland, School of Mathematics and Physics, Brisbane, Qld 4072, Australia}
\affiliation{University Paris Sud, CNRS, LPTMS, UMR 8626, Orsay 91405, France}
\author{Sebastian W\"{u}ster}
\affiliation{The University of Queensland, School of Mathematics and Physics, Brisbane, Qld 4072, Australia}
\affiliation{Max Planck Institute for the Physics of Complex Systems, N\"{o}thnitzer Strasse 38, 01187 Dresden, Germany}
\author{Christopher J. Vale}
\affiliation{The University of Queensland, School of Mathematics and Physics, Brisbane, Qld 4072, Australia}
\affiliation{ARC Centre of Excellence for Quantum-Atom Optics and Centre for Atom Optics and Ultrafast Spectroscopy, Swinburne University of Technology, Melbourne, Vic 3122, Australia}
\author{Norman R. Heckenberg}
\author{Halina Rubinsztein-Dunlop}
\author{C. A. Holmes}
\author{G. J. Milburn}
\author{Matthew J. Davis}
\email{mdavis@physics.uq.edu.au}
\affiliation{The University of Queensland, School of Mathematics and Physics, Brisbane, Qld 4072, Australia}

\begin{abstract}
The study of dynamical tunnelling in a periodically driven anharmonic potential probes the quantum-classical transition via the experimental control of the effective Planck's constant for the system.
In this paper we consider the prospects  for observing dynamical tunnelling with
ultracold atoms in magnetic microtraps on atom chips.  We outline the driven anharmonic potentials that are possible using standard magnetic traps, and find the Floquet spectrum for one of these as a function of the potential strength, modulation, and effective Planck's constant.  We develop an integrable approximation to the non-integrable Hamiltonian and find that it can explain the behaviour of the tunnelling rate as a function of the effective Planck's constant in the regular region of parameter space.  In the chaotic region we compare our results with the predictions of models that describe chaos-assisted tunnelling.  Finally we examine the practicality of performing these experiments in the laboratory with Bose-Einstein condensates.

\end{abstract}

\maketitle

\section{Introduction}

Tunnelling is one of the ubiquitous features of wave mechanics. 
A familiar example of  quantum tunnelling of a single particle occurs in a
time-independent spatially symmetric double-well potential. A quantum particle
initially located in one of the wells with an energy below the maximum of the
potential barrier between the  wells will tunnel between them, despite this
being classically forbidden (see, for instance, Ref.~\cite{Cohen}).  However,
tunnelling is a more general phenomenon that can be observed in situations
involving other types of symmetries and barriers.

In this paper we are interested in dynamical tunnelling in classically non-integrable systems, i.e. Hamiltonian systems with more degrees of freedom than constants of motion. The {classical} dynamics of such systems is known to exhibit chaotic features \cite{Lichtenberg}. However, for a Hamiltonian of the form $H_0+\epsilon V$ where $H_0$ is integrable and $\epsilon V$ is a sufficiently small chaos-inducing perturbation (a quasi-integrable system), some constants of motion are locally conserved leading to the formation of so-called ``Kolmogorov-Arnol'd-Moser (KAM) tori'' \cite{Lichtenberg}.  Within these tori the dynamics of the system is still regular. Poincar\'e surfaces of sections of such systems show characteristic patterns of seas of chaos surrounding islands of regular motion [see for instance Fig.~\ref{fig-Poinc}(d)].

While particles cannot classically escape {the regions bound by the} KAM tori, quantum mechanical particles are able to tunnel through KAM barriers to symmetry-related islands \cite{Reichl}. Due to the similarities with spatial tunnelling in double-wells and the dynamical origin of the barrier the quantum particle crosses, this phenomenon was named \emph{dynamical tunnelling} by Davis and Heller~\cite{Davis1981}.

Laser-cooled atomic gases confined in magneto-optical traps have proven to be useful  systems for the demonstration of single-particle matter wave phenomena.  Impressive examples include Kapitzsa-Dirac scattering~\cite{PhysRevLett.56.827}, Bragg scattering~\cite{PhysRevLett.60.515}, two-slit interference~\cite{PhysRevLett.77.4}, and Wannier-Stark ladders~\cite{Wilkinson:wannier:stark}.  One of the first experimental demonstrations of dynamical tunnelling by Steck \etal\ began with velocity-selected cold atoms from a magneto-optical trap~\cite{Steck2001}.

Further decreasing the temperature of cold atom systems via evaporative cooling can lead to the formation of a Bose-Einstein condensate (BEC) --- a ``giant'' matter wave.  This can lead to practical advantages in studying the physics of matter waves, as the  increase in phase space density combined with  macroscopic coherence allows for the single shot visualisation of single-particle matter wave phenomena, such as interference \cite{Andrews1997b} and quantum tunnelling \cite{Albiez2005a}.  However, the repulsive interactions common in atomic BEC can cause additional complications.  A second experimental demonstration of dynamical tunnelling by Hensinger \etal\ used a Bose-Einstein condensate as their starting point~\cite{Hensinger2001}.    The BEC was released from its initial trap, and allowed to expand until the interaction energy was negligible.  Single-particle dynamical tunnelling was observed in the following dynamics~\cite{PhysRevA.70.013408}. Another recent experiment on dynamical tunnelling  succeeded in extracting the full phase-space representation of the quantum state of a kicked top~\cite{Chaudhury2009,Steck2009}.

A common theoretical model for dynamical tunnelling is that of the nonlinear
pendulum, with the dimensionless Hamiltonian
\begin{eqnarray}
H = \frac{p^2}{2} + \kappa (1 + 2 \epsilon \cos t ) \sin^2\left(\frac{q}{2}\right),
\label{eqn:Hstanding}
\end{eqnarray}
{where $p$, $q$, and $t$ are the dimensionless momentum, position, and time variables respectively, $\kappa$ is the potential strength, and $\epsilon$ is the strength of the driving. }  This potential can be realised experimentally with cold atoms placed in an intensity
modulated optical standing wave, and this was the system realised by  Steck \etal~\cite{Steck2001}  and  Hensinger \etal~\cite{Hensinger2001}. 

One of the motivations for the theoretical study of dynamical tunnelling has been to explore the
boundary between classical and quantum dynamics. The relevant parameter is the
effective Planck's constant, defined by the commutator $[\hat{q},\hat{p}] = i
\kbar$ of the dimensionless position $q$ and momentum $p$ variables of the Hamiltonian,
Eq.~(\ref{eqn:Hstanding}).  By varying experimental parameters it is in principle
feasible to carry out experiments that are identical apart from differing
values of $\kbar$.  This would, for example, lead to a variation in the rate of
tunnelling, and this has been studied in a large number of theoretical papers \cite{Ullmo1996,Mouchet2003,Mouchet2001,Brodier2001,Brodier2002,Eltschka2005,Tomsovic1994,Leyvraz1996,Podolskiy2003,Mouchet2006,Baecker2008}. However, the  experiments performed by Steck \etal~\cite{Steck2001}  and  Hensinger \etal~\cite{Hensinger2001}  considered single,  large values of $\kbar$, and while impressive, they did not probe the quantum-classical transition that occurs as $\kbar \rightarrow 0$.  Here we revisit the problem of observing dynamical tunnelling in a experiment with ultracold atoms for a range of effective Planck's constant $\kbar$ to probe the quantum-classical transition.  

As well as working at large $\kbar$, the  modulated standing-wave potential for cold atoms used in Refs.~\cite{Steck2001,Hensinger2001} to study dynamical tunnelling does not directly correspond to the classical non-linear pendulum. The phase space of the pendulum has periodic boundary conditions for the position variable, while atoms  in an optical standing wave can move between lattice sites.  The value of $\kbar$ used by Hensinger \emph{et al.}~\cite{Hensinger2001} was
sufficiently large that during the dynamics the atoms were not confined to a single well of the standing wave~\cite{PhysRevA.70.013408,Hensinger2001,Upcroft}. Thus, the Husimi
functions of the tunnelling Floquet states were not confined to a single lattice site, and long range coherence played an important role in the observed dynamics~\cite{PhysRevA.70.013408,Hensinger2001,Upcroft}.  This complicates the picture that was presented in Ref.~\cite{Hensinger2001} of a classical particle tunnelling to an oscillatory mode with the same amplitude but 180$^\circ$ out of phase.  One way of avoiding this would be to use a trapping potential with a
single minimum rather than multiple minima. Then, within the approximation that
all atoms share the same {single-particle} wave function, all atoms will experience the same
dynamics. 

For a one dimensional system, a sinusoidal modulation of the trapping potential can lead to a mixed phase space with symmetric islands of regular motion. If the potential is anharmonic then dynamical tunnelling can occur between the period-one islands in the phase space. However, to a first approximation almost all ultracold atom
experiments occur in harmonic trapping potentials.  One possible realisation of an anharmonic potential for ultracold atoms is the radial trapping potential formed by the magnetic field from current-carrying microscopic
wires combined with a homogeneous bias field on an atom chip \cite{Fortagh2007a}. While such potentials are harmonic at the centre,
beyond a certain length scale they become linear. Also, these microtraps can be made
very tight, potentially giving access to a large range of $\kbar$, a crucial prerequisite for studies of the quantum-classical transition and so far  not achieved in cold atom experiments.

While quantum chaos and dynamical tunnelling are purely single-particle effects, the preparation of a BEC greatly simplifies the efficient loading of small regions of phase space. Therefore, in this paper we consider the possibilities for studies of quantum chaos using BECs confined by magnetic potentials on an atom chip. As long as the condensate
is sufficiently dilute, its dynamical tunnelling is to a good approximation dominated by single-particle physics, to which we  devote a large part of this article.
We outline our model in \sref{sec:model}, before describing our numerical results for dynamical tunnelling in a 1D atom chip potential in \sref{sec:results}. We analyse these results using different theoretical methods, and in particular derive an integrable approximation that explains some of the observed features.  In \sref{sec:exp} we consider the practicalities of realising these experiments, including the effect of mean-field interactions in the BEC, before concluding in \sref{sec:concl}. In \aref{sec:trap} we outline the derivation of the possible potentials realizable with an atom chip, essentially consisting of an infinite wire carrying a time-dependent current combined with a time-dependent bias magnetic field. In \aref{sec:gpe} we derive the reduced dimensional Gross-Pitaevskii equation that we use for simulations in  \sref{sec:exp}.

\section{Model}\label{sec:model}

\subsection{Hamiltonian}
We consider a time-dependent potential realised by a current-carrying wire and a homogeneous bias magnetic field.  Reducing the problem to one dimension and making use of dimensionless units (see \aref{sec:trap}), the following dimensionless Hamiltonian can be realised:
\begin{eqnarray}
H_m &=& \frac{p^2}{2} + \kappa (1+ \epsilon \cos t)(1 +   q^2)^{1/2}.
\label{eqn:H1text}
\end{eqnarray}
{Here the dimensionless momentum $p$, position $q$, and time $t$ variables can be converted to SI units by multiplying by the quantities
\begin{eqnarray}
\label{dimlessunitssscales}
\bar{p} = \frac{B_0 m \Omega}{B'}, \quad 
\bar{q} = \frac{B_0}{B'}, \quad \bar{t} = \frac{1}{\Omega},
\end{eqnarray}
respectively.}
Here $B'$ is the magnetic field gradient at the trap centre, $B_0$ is the magnetic field offset, $\Omega$ is the frequency of modulation of the trapping potential, and $m$ is the mass of the atom.  The amplitude of the driving potential is given by $\epsilon$, and
the strength of the potential is
\begin{equation}
\kappa = \frac{g m_F \mu_B B'^2}{m B_0 \Omega^2} \equiv
\frac{\omega_r^2}{\Omega^2},
\end{equation}
where $g$ is the Land\'e $g$-factor, $m_F$ identifies the magnetic sublevel of the atom, $\mu_B$ is the Bohr magneton, and $\omega_r$ is the harmonic trapping frequency about the {minimum} of the potential for small amplitude {oscillations}.
A detailed derivation of the Hamiltonian Eq.~(\ref{eqn:H1text}) can be found in {\aref{sec:trap}.}

\subsection{Classical dynamics}

\begin{figure}[t]
\begin{center}
\includegraphics[width=8.6cm]{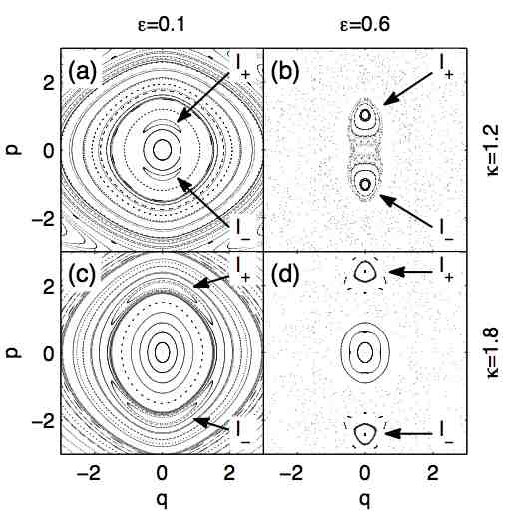}
\caption{Poincar\'e sections for the {classical dynamics of the} atom-chip Hamiltonian, Eq.~(\ref{eqn:H1text}) {at $t=0$}, illustrating the influence of the potential strength $\kappa$ and modulation amplitude $\epsilon$ on the classical dynamics. {We show results for the parameters:} (a) $\kappa = 1.2$, $\epsilon=0.1$. (b) $\kappa = 1.2$, $\epsilon=0.6$.
(c) $\kappa = 1.8$, $\epsilon=0.1$. (d) $\kappa = 1.8$, $\epsilon=0.6$. }
\label{fig-Poinc}
\end{center}
\end{figure}

The parameters $\kappa$ and $\epsilon$  are experimentally tunable, and altering them allows the investigation of various
dynamical regimes.  Example Poincar\'e sections for two values of $\kappa$ and $\epsilon$ are shown in Fig.~\ref{fig-Poinc}. 

\begin{figure}[t]
\begin{center}
\includegraphics[width=6cm]{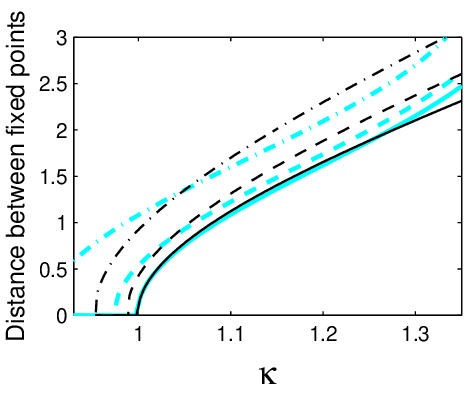}
\caption{(Color online)
Distance in phase-space between the period-one resonances at $t = 0$ as a function of potential strength $\kappa$ for three values of modulation strength $\epsilon$.  The black curves are determined from the Poincar\'e section for the full Hamiltonian Eq.~(\protect\ref{eqn:H1text}).
The  cyan (gray) curves are the corresponding result derived from the integrable approximation to the full Hamiltonian, Eq.~(\protect\ref{eqn:Hi}), as introduced in Sec.~\ref{sec:analysis-QRR}.  {These give} an indication of the validity of the integrable approximation. Solid lines: $\epsilon = 0.1$. Dashed lines: $\epsilon = 0.4$. Dash-dot lines: $\epsilon = 0.8.$}
\label{fig-distbetweenres}
\end{center}
\end{figure}

In this article we consider the dynamics associated with the fixed points resulting from the 1:1 resonance of the unperturbed motion (described by $H_0=p^2/2+\kappa\sqrt{1+q^2}$) with the period of the perturbation $\epsilon V=\epsilon\kappa\cos t\sqrt{1+q^2}$. As the quantity $\kappa$ is increased {from zero}, two such fixed points appear at the origin of the phase-space at $\kappa=1$ \footnote{Note that this value is exact only in the limit of an unperturbed system.} and then move away from each other,  as plotted in Fig.~\ref{fig-distbetweenres}.

As $\epsilon$ is increased, the KAM tori constraining the motion to regular behaviour are progressively destroyed, and larger regions of chaos appear about the islands centered on the  period-one resonances.


\subsection{Quantum dynamics}
The effective Planck's constant for this system is the commutator of the dimensionless quantum position and momentum operators
\begin{equation}
\kbar = i [\hat{p},\hat{q}] = \frac{\hbar B'^2}{B_0^2 m \Omega}.
\end{equation}
Larger  values of $\kbar$ correspond to systems that are ``more quantum''.  It is possible to experimentally tune this parameter while retaining fixed values of $\kappa$ and $\epsilon$.

In a quantum system, dynamical tunnelling will take place between the period-one islands of regular motion $\cal{I}_+$ and $\cal{I}_-$ of the classical Hamiltonian because of their time-reversal symmetry. This can be understood {using} Floquet theory, a formalism used to describe systems with a periodic time dependence \cite{Mouchet2003, Mouchet2001, Luter2002, Brodier2001, Brodier2002, Averbukh2002, Hensinger2001, Steck2001, Eltschka2005, Grossmann1991, Reichl}. The  Floquet operator $\hat{F}$ describes quantum evolution for one period of the potential modulation, $T_0$, and its eigenstates are invariant under a {$T_0 = 2 \pi \bar{t}$} time translation.  These are similar to the eigenstates of a time-independent Hamiltonian, which are invariant under any time translation. In our system, the tunnelling states are even and odd {superpositions} of states localized on the islands ${\cal I}_+$ and ${\cal I}_-$.  These are expected to be eigenstates of the system's Floquet operator $F_m$. As noted in the introduction, $|\psi_{\rm even}\rangle$ and $|\psi_{\rm odd}\rangle$ are analogous to the ground state and first excited state of a double-well system respectively. Because the Floquet operator is unitary, it has eigenvalues of the form
\begin{equation}
f=\exp\left(-2\pi i {E} / \kbar\right),
\end{equation} where ${E}$ is the corresponding eigenstate's \emph{quasi-energy} --- the generalization of the notion of energy to time-periodic systems.

\begin{figure}
\begin{center}
\includegraphics{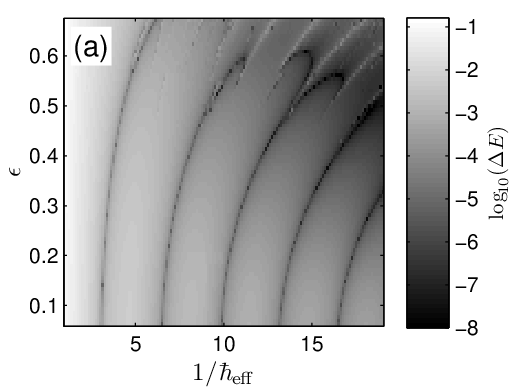}
\includegraphics{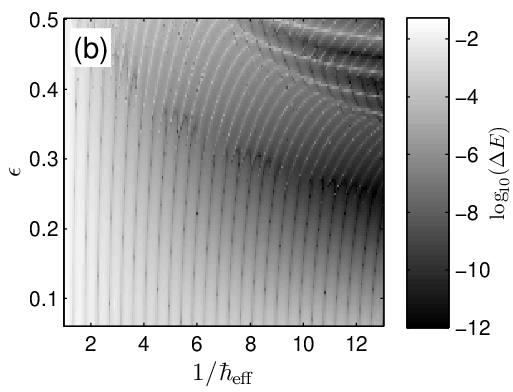}
\caption{Quasi-energy splitting $\Delta{E}$ of the tunnelling Floquet states
 as a function of the inverse effective Planck's constant $1/\protect\kbar$ and the modulation strength $\epsilon$ for the atom-chip Hamiltonian~Eq.~(\ref{eqn:H1text}) for two different potential strengths.
 (a) $\kappa = 1.2$; (b) $\kappa = 2.0$.
}
\label{fig-graph2ds}
\end{center}
\end{figure}
\begin{figure}
\begin{center}
\includegraphics{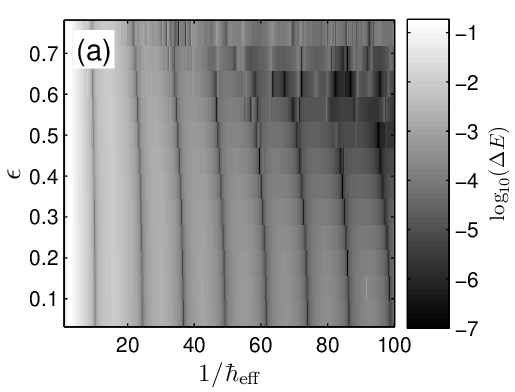}
\includegraphics{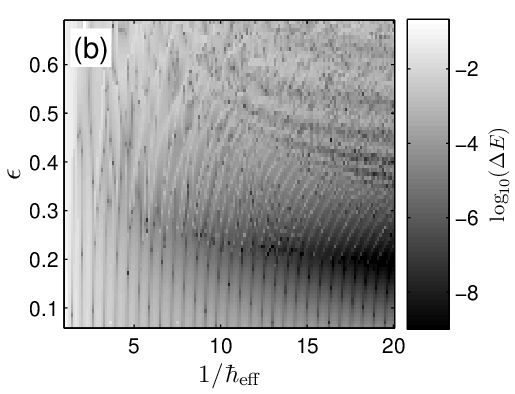}
\caption{Quasi-energy splitting $\Delta{E}$ of the tunnelling Floquet states
 as a function of the inverse effective Planck's constant $1/\protect\kbar$ and the modulation strength $\epsilon$ for the
 (a) Quartic oscillator {with Hamiltonian given by Eq.~(\ref{eqn:quartic}) and} $\kappa=2$; (b) Nonlinear pendulum {with Hamiltonian given by Eq.~(\ref{eqn:nlpend}) and} $\kappa = 1.5$.
}
\label{fig-otherhams}
\end{center}
\end{figure}

\subsection{Numerical solution of the Floquet spectrum}
\label{sec:num_results}
To study dynamical tunnelling in this system, we must numerically determine the Floquet spectrum and identify the even and odd Floquet states that are localised on the period-one fixed points of the Poincar\'e section.  {These will be dependent on} the potential strength $\kappa$, modulation amplitude $\epsilon$ and effective Planck's constant $\kbar$.   {As the trapping potential is symmetric, the even and odd Floquet states are uncoupled.}  Our numerical procedure is to choose a {regular grid in position space, and then} separately evolve a basis of even and odd combinations of {position eigenfunctions} for one period of the modulation. This determines the Floquet matrix $F_m$ for the Hamiltonian $H_m$, which we diagonalize to find the Floquet states.
{The tunnelling states are identified as being the even (odd)  Floquet states having the largest overlaps with an even (odd) superposition of coherent states centered on the period-one fixed points of the classical phase space.}
We can therefore expect a localised state on one island to be formed by $|\phi_+ \rangle = ( |\psi_{\rm even}\rangle + |\psi_{\rm odd}\rangle)/\sqrt2$, {and this will tunnel to a state $|\phi_- \rangle = (|\psi_{\rm even}\rangle - |\psi_{\rm odd}\rangle)/\sqrt{2}$ localised on the opposite island. } The corresponding tunnelling period is given by $T_{\rm tunnel}=2\pi\kbar/\Delta{E}$, where $\Delta{E}$ is the quasi-energy splitting of the tunnelling states: $\Delta{E}=|{E}_{\rm even}-{E}_{\rm odd}|$.

\section{Results}
\label{sec:results}
In this section we investigate the dependence of the quasi-energy splitting $\Delta{E}$ on the tunable parameters for our system $\kappa$, $\epsilon$, and $\kbar$. Typical results are shown in Fig.~\ref{fig-graph2ds} for (a) $\kappa = 1.2$ and (b) $\kappa = 2.0$.
These figures have several noteworthy features, many of which {have been observed previously} in other systems \cite{Mouchet2003, Mouchet2001, Tomsovic1994, Grossmann1991, Averbukh2002}. 

\subsection{Overview}

For our analysis, we divide the parameter space shown in Fig.~\ref{fig-graph2ds}  into two regions according to the characteristic behavior of the quasi-energy splitting.
 Firstly, we see in the bottom-left corner of \fref{fig-graph2ds}(a--b) that for a large parameter range the behaviour of $\Delta{E}$ is not completely unruly, but instead some grooves and plateau structures are obvious. The dependence of $\Delta{E}$ on $1/\kbar$ differs from the exponential behaviour characteristic of Hamiltonians of the form $H=p^2/2+V(q)$, while still being quite smooth. In this regime, which we call quantum regular regime (QRR), the quasi-energy splitting  periodically falls to zero. To the best of our knowledge, this has not been commented on in detail in previous studies --- we account for the origin of these grooved structures in \sref{sec:analysis-QRR}. This result constitutes the most significnate theoretical development of this paper.

Secondly, in the top right corners of Fig.~\ref{fig-graph2ds}(a--b), classical chaos has the strongest influence on the system. This regime is characterised by the dramatic and apparently disorderly fluctuations of $\Delta{E}$ \cite{Mouchet2003,Mouchet2001,Leyvraz1996}, and we refer to it as  classical chaotic regime (CCR).  In our discussion of the CRR, \sref{sec:analysis-CCR}, we apply existing theories by Podolskiy and Narimanov \cite{Podolskiy2003} and Eltschka and Schlagheck \cite{Eltschka2005} to our specific scenario. Comparison of these theories with our numerical simulations shows good agreement and completes our analysis of the present atom chip system in the single-particle regime.

We note that the division described above is not peculiar to the atom-chip Hamiltonian, but  appears to be a general feature of driven quasi-integrable systems.  We have also diagonalized the Floquet operator for other nonlinear potentials such as the quartic oscillator
\begin{equation}
H=\frac{p^2}{2}+\kappa(1+\epsilon\cos t)q^4,
\label{eqn:quartic}
\end{equation}
as well as the nonlinear pendulum
\begin{equation}
H=\frac{p^2}{2}- \kappa(1+\epsilon\cos t)\cos q,
\label{eqn:nlpend}
\end{equation}
and found the same features.   Sample results are shown in Fig.~\ref{fig-otherhams}. The latter Hamiltonian has been studied elsewhere both theoretically \protect\cite{Mouchet2003,Mouchet2001,Luter2002,Averbukh2002} and experimentally \protect\cite{Steck2001,Hensinger2001}.

\subsection{Quantum regular regime}
\label{sec:analysis-QRR}
\begin{figure*}
\begin{center}
\includegraphics[width=17.2cm]{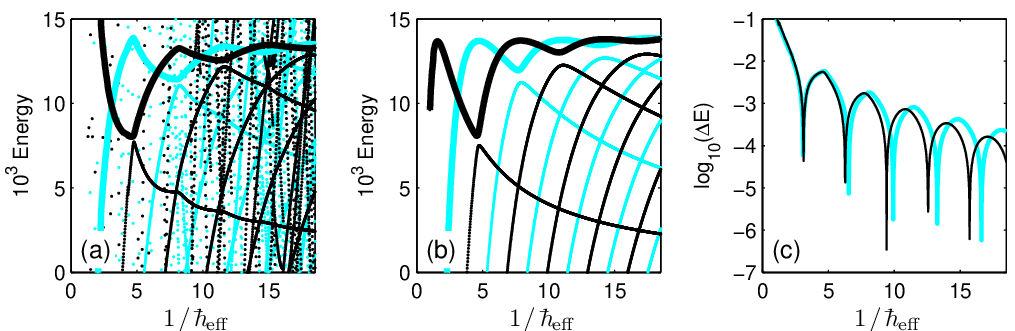}
\caption{(Color online) Comparison of the Floquet spectrum of the full Hamiltonian $H_m$, Eq.~(\protect\ref{eqn:H1text}), and the energy spectrum of the
integrable approximation $H_i$, Eq.~(\protect\ref{eqn:Hi}), for $\kappa=1.2$ and $\epsilon=0.1$.
(a) The quasi-energy spectrum of the full Hamiltonian as a function of $1/\protect\kbar$.  The black points are for the even Floquet states, and the cyan (gray) points are for the odd Floquet states.  The points corresponding to the tunnelling states are joined with thick solid lines of the same color.
(b) The energy spectrum of the integrable Hamiltonian as a function of $1/\protect\kbar$, with the same color scheme as for (a).  The structures here are very similar to those in (a), indicating that the integrable approximation captures the important features of the full Hamiltonian in this regime.
(c) A plot of the quasi-energy difference of the even and odd tunnelling states as a function of $1/\protect\kbar$. Black curve: full Hamiltonian.  Cyan (gray) curve: integrable approximation.
}
\label{fig-espectrum}
\end{center}
\end{figure*}

We define the quantum regular regime as the region in parameter space where the quasi-energy splitting $\Delta{E}$ varies smoothly, apart from the apparently periodic troughs where the splitting goes to zero.  In both 
 Fig.~\ref{fig-graph2ds}(a) and (b) it is the region for which $\epsilon \alt 0.3$.
The origin of this behaviour
can be seen by plotting the quasi-energy spectrum of the Floquet operator $F_m$ as in Fig.~\ref{fig-espectrum}(a--b). We find that at regularly-spaced values of $1/\kbar$, a Floquet state not involved in the tunnelling  deflects  the tunnelling state of the same parity through an avoided crossing scheme very similar to the result of a typical first-order perturbation theory calculation.

\begin{figure}
\begin{center}
\resizebox{8.6cm}{!}{\includegraphics{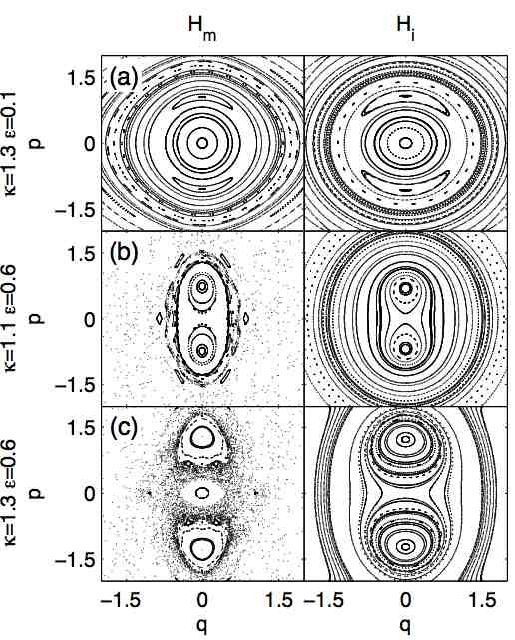}}
\caption{Comparison of the Poincar\'e sections for the full Hamiltonian $H_m$, Eq.~(\ref{eqn:H1text}), and the integrable approximation $H_i$, Eq.~(\ref{eqn:Hi}) {at $t=0$}, for (a) $\kappa=1.3$ and $\epsilon=0.1$, (b) $\kappa=1.1$ and $\epsilon=0.6$, (c) $\kappa=1.3$ and $\epsilon=0.6$. Although the presence of chaos for $H_m$ in this latter regime makes the  Poincar\'e sections of $H_m$ and $H_i$ look rather different, we find good agreement for the regular regions. The quantum behaviour of the two systems agree for relatively large values of $\protect\kbar$ as shown in \sref{sec:analysis-QRR}.
}
\label{fig-comparePoinc}
\end{center}
\end{figure}

This scenario has previously been invoked in the context of tunnelling or dynamical tunnelling suppression \cite{Grossmann1991,Averbukh2002}, often in order to describe the phenomena inducing the CCR fluctuations \cite{Mouchet2003,Mouchet2001,Holder2005}. More generally, 
{it has been noted that level repulsion occurs for dynamical tunnelling in the presence of a chaotic sea}
 \cite{Leyvraz1996,Podolskiy2003,Eltschka2005}. However, we demonstrate here that this effect has no chaotic origin. This conclusion was also reached in \cite{Holder2005} by different means. Here
we make use of secular perturbation theory \cite{Lichtenberg}, and
assume that $\kappa-1$ is of order $\epsilon$ and therefore  within the islands of regular motion we have both $q={\cal O}(\sqrt{\epsilon})$ and $p={\cal O}(\sqrt{\epsilon})$.
This allows us to
 to derive a second-order integrable approximation $H_i$ to the full atom-chip Hamiltonian  $H_m$ given by Eq.~(\ref{eqn:H1text}) as  \cite{Lichtenberg}
\begin{widetext}
\begin{equation}
H_i=\frac{3}{1024}\left(p^2+q^2\right)^3+
	\frac{3(\kappa-2)}{64}\left(p^2+q^2\right)^2+
	\frac{6(\kappa-1)-3(\kappa-1)^2+\epsilon^2}{24}p^2+
	\frac{6(\kappa-1)-3(\kappa-1)^2-5\epsilon^2}{24}q^2 \textrm{.}
\label{eqn:Hi}
\end{equation}
\end{widetext}
We find that this integrable approximation represents the classical dynamics of $H_m$ reasonably well up to $\kappa=1.3$ and $\epsilon=0.6$.  This can be seen by comparing the Poincar\'e sections for the full Hamiltonian~Eq.~(\ref{eqn:H1text}) and the integrable approximation Eq.~(\ref{eqn:Hi}) in Fig.~\ref{fig-comparePoinc}.  Also, the prediction of the distance between the period-one fixed points of the integrable approximation  Eq.~(\ref{eqn:Hi}) and the full Hamiltonian~Eq.~(\ref{eqn:H1text}) is shown in Fig.~\ref{fig-distbetweenres} as a function of $\kappa$ for three values of $\epsilon$.

To proceed we quantize the integrable Hamiltonian $H_i$, Eq.~(\ref{eqn:Hi}).  In principle we could choose any operator ordering; given the symmetry of Eq.~(\ref{eqn:Hi}) we choose symmetric ordering.  This is then easily diagonalised numerically, and the {resulting} spectrum is plotted in  Fig.~\ref{fig-espectrum}(b).  This should be compared to the spectrum of the  Floquet operator $F_m$ for the full Hamiltonian shown in Fig.~\ref{fig-espectrum}(a).  Qualitatively the spectra are similar and quantitatively the prediction for the quasi-energy splittings of the tunnelling states are in good agreement with that found from the full Hamiltonian as shown in Fig.~\ref{fig-espectrum}(c).  We conclude that the tunnelling states of the integrable approximation contain the essential features of the full system for {$\kappa \alt 1.3$, $\epsilon \alt 0.6$, and $1/\kbar \agt 1$. It was previously shown that the enhancement of tunnelling does not require chaos but can originate from  avoided crossings in the Floquet spectrum~\cite{Holder2005}. Here we have demonstrated that even non-integrability is not necessary.}

\begin{figure}
\begin{center}
\includegraphics{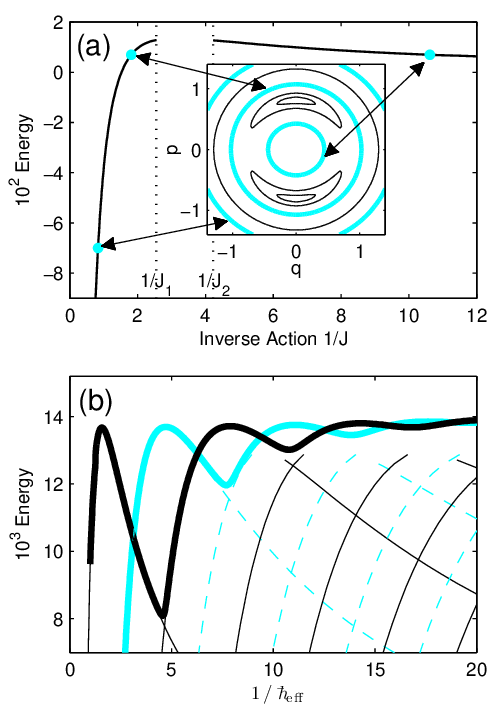}
\caption{(Color online) (a) Energy versus inverse action for the trajectories of the integrable approximation $H_i$ that are not located on the islands (solid black curve).  The gap occurs due to the {presence of the islands}. The inset shows selected trajectories of $H_i$.  The energy of the trajectories in cyan (gray) are denoted in the main figure by cyan (gray) points as indicated by the arrows. (b) Energy of the semi-classical and quantum states as a function of $1/\protect\kbar$. {The thin curves are the EBK quantization of the trajectories of $H_i$, Eq.~(\ref{eqn:Hi}). The thin solid black curves correspond to even parity states, and the thin dashed cyan (gray) curves to odd parity states.  The thick black curve is the energy of the even tunnelling state, and the thick cyan (gray) curve is the energy of the odd tunnelling state, both found via the numerical diagonalization of $H_i$, Eq.~(\ref{eqn:Hi}).} The avoided crossings of the quantum tunnelling states with other states of the same parity are clearly visible.}
\label{fig-enaction}
\end{center}
\end{figure}

To understand the {quasi-periodic} vanishing of the difference in quasi-energy of the tunnelling Floquet states as a function of $\kbar$, we first plot in Fig~\ref{fig-enaction}(a) the energy of the classical trajectories of $H_i$ not located on the islands for $\kappa = 1.2$ and $\epsilon = 0.1$ in Fig.~\ref{fig-enaction}(a) as a function of the inverse action $J^{-1}$, where \cite{Lichtenberg}
\begin{equation}
J = \frac{1}{2\pi}\oint p\,dq.
\end{equation}
 { A gap exists in the curve for the range of the inverse action $2.5 \alt J^{-1} \alt 4.2$  corresponding to energies ${E} \agt 1.29 \times 10^{-2}$ for which the classical trajectories are instead located on the islands.}

We can now apply the Einstein-Brillouin-Keller (EBK) quantization method, which states that in order to correspond to a quantum eigenstate, the trajectories must satisfy
\begin{equation}
J=\kbar(n+1/2)~\textrm{with}~n\in\mathbb{N}.
\end{equation}
This will discretise the curve {of energy versus inverse action} in Fig.~\ref{fig-enaction}(a) for any particular value of $\kbar$.  Now as $1/\kbar$ is {increased}, the spacing of the points along the $x$-axis will {also be increased}, and particular eigenstates will move from the left-hand part of the curve (outside of the islands) to the right-hand part (inside the islands), and have a strong effect on the tunnelling states.  The energy of EBK quantized trajectories for $n=0$ to $n=10$ is plotted in
 Fig.~\ref{fig-enaction}(b) --- even trajectories as thin black solid curves and odd trajectories as thin cyan (gray) dashed curves.  The thick solid black curve is the energy of the even tunnelling state, and the thick solid cyan (gray) curve is the odd tunnelling state, {both} found from numerical diagonalisation of the integrable approximation to the full Hamiltonian, Eq.~(\ref{eqn:Hi}). In Fig.~\ref{fig-espectrum} it is easy to see the avoided level crossings between Floquet states with the same parity.  The curve for the quasi-energy of each tunnelling Floquet state  changes direction at each avoided crossing, resulting in regular crossings of the quasi-energies of the odd and even Floquet tunnelling states, resulting in $\Delta E =0$ and an infinite tunnelling period.  It seems quite reasonable that this explanation applies directly to the quasi-integrable Hamiltonian (\ref{eqn:H1text}) in the QRR regime.

We have found that the spacing of the occurrences of $\Delta E = 0$ as a function of $1/\kbar$ decreases with increasing $\kappa$. For example, it can be seen in Fig.~\ref{fig-graph2ds} that the spacing of the valleys as a function of $1/\kbar$ decreases between (a) with $\kappa = 1.2$ and (b) with $\kappa = 2.0$. This can be explained in the following manner. We find that that the period-one islands remain of a similar phase space area for a given $\kappa$.  Let  the \emph{minimum} action of any {closed} trajectory that lies entirely \emph{outside} the islands be denoted as $J_1$.  Likewise,  let the \emph{maximum} action of any {closed} trajectory that lies entirely \emph{inside} the islands be denoted as $J_2$.
Then the $n$th (with $n$ large enough) crossing will occur between $1/\kbar=(n+3/2)/J_1$ and $(n+1/2)/J_2$.
As $\kappa$ increases, the distance between the islands in phase space increases, meaning $J_1$ and $J_2$ increase, and the spacing between the $n$th and $(n+1)$th quasi-energy degeneracy of the tunnelling states will decrease.  This is in agreement with the behavior of the Floquet spectrum of the full system.

In summary, in this section we have shown that the quantised integrable approximation \eref{eqn:Hi}  accounts for \emph{all} the significant features  of dynamical tunnelling of the full system Hamiltonian \eref{eqn:H1text} in the QRR.  Therefore we have demonstrated that the underlying classical non-integrability of the full system has little effect in the QRR.

\subsection{Classical chaotic regime}
\label{sec:analysis-CCR}
%
%

The defining characteristic of the classical chaotic regime is the irregular and large fluctuations of the quasi-energy splitting $\Delta{E}$ as the parameters $1/\kbar$ and $\epsilon$ are varied. Before considering the fluctuations, we comment on the coarse behaviour of the tunnel splitting as we enter the CCR from the QRR, by choosing $\epsilon$ near the upper boundaries of Fig.~\ref{fig-graph2ds}(a--b) and then increasing $1/\kbar$.

\begin{figure}
\begin{center}
\includegraphics[width=8.6cm]{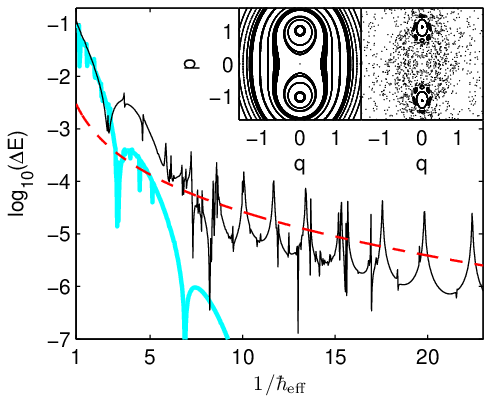}
\caption{(Color online) Quasi-energy splitting of the tunnelling Floquet states versus $1/\protect\kbar$ for $\kappa=1.2$, $\epsilon=0.75$.  Thin solid black line: exact result. Thick cyan (gray) solid line: numerical diagonalization of the integrable approximation.  Dashed red line: result of the Podolskiy-Narimanov theory with the proportionality coefficient of Eq.~(\ref{eq-PN}) as a fitting parameter. Inset: Poincar\'e sections of the full Hamiltonian (right) and of its integrable approximation (left). The tunnelling behaviour of the full Hamiltonian $H_m$ [Eq.~(\ref{eqn:H1text})]  agrees with the integrable approximation $H_i$ [Eq.~(\ref{eqn:Hi})] as long as the quantum coarse-graining effect prevents the quantum particle from seeing the layers of chaos. For smaller values of $\protect\kbar$, tunnelling is enhanced by the presence of chaos in the underlying classical phase-space according to the Podolskiy-Narimanov model \cite{Podolskiy2003}.}
\label{fig-transition}
\end{center}
\end{figure}

\begin{figure}
\begin{center}
\includegraphics[width=8.6cm]{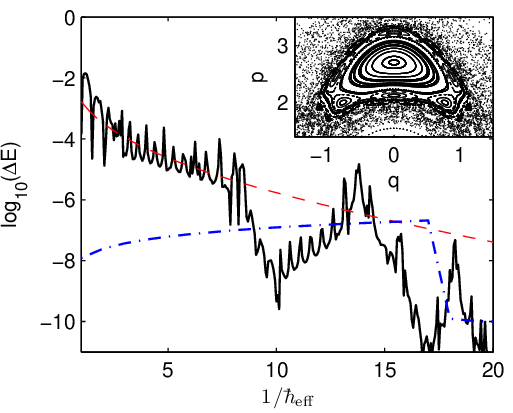}
\caption{(Color online) Quasi-energy splitting of the tunnelling Floquet states as a function of $\protect\kbar$ for $\kappa=2$, $\epsilon=15/32$.  Solid black line: exact numerical result.  Red dashed line: Podolskiy-Narimanov theory (one fitting parameter).  Blue dash-dot line: Eltschka-Schlagheck theory (no fitting parameter). Inset: Poincar\'e section centered on $\cal{I}_+$. One secondary resonances chain is clearly dominant. For large values of $\protect\kbar$, the quantum coarse-graining prevents the system from seeing these small phase-space structures, and its behaviour is then well described by Eq.~(\protect\ref{eq-PN}). For smaller values of $\protect\kbar$ the Eltschka-Schlagheck model  \cite{Eltschka2005}, which takes the island chain into account, shows much better agreement with the numerical data.}
\label{fig-kbargraph}
\end{center}
\end{figure}

Averaging over the fluctuations of the quasi-energy difference of the Floquet states as a function of $1/\kbar$, {we find} that the overall tendency of the tunnelling splitting variation resembles an exponential decrease with increasing $1/\kbar$, {as expected for dynamical  tunnelling phenomena} \cite{Tomsovic1994,Wilkinson1987}.  It is not surprising that the tunnel splitting tends to decrease as the {potential strength $\kappa$ is increased and the two islands move away from each other in phase-space.}  However, it should be noted that after initially decreasing with increasing $\epsilon$, $\Delta{E}$ starts increasing again as the system enters the CCR \cite{Mouchet2003}. This latter phenomenon can be attributed to so-called ``chaos-assisted tunnelling''. This occurs when the dominant transport mechanism from one island to the other is no longer direct quantum transport as in a regular system, but instead a three-step process.  Firstly particles tunnel from the centre of one island to the sea of chaos, then are \emph{classically} transported through the chaotic sea from the vicinity of this island to the vicinity of the other, and then eventually tunnel from the sea of chaos to the center of this latter island \cite{Podolskiy2003}. The complete behavior of $\Delta{E}$ as $1/\kbar$ is increased is shown in Figs.~\ref{fig-transition} and \ref{fig-kbargraph}. A number of regions with distinct changes in the average slope of the $\log_{10}(\Delta{E})$ versus $1/\kbar$ curve are evident. These can indeed be partially accounted for by changes of the dominant transport mechanism from one island to the other. 

For large values of $\kbar$, we expect the quantum behaviour of chaotic systems to approach that of regular systems, as the quantum ``coarse-graining'' makes them insensitive to the presence of fine phase-space structures. We are hence in the QRR, where the quasi-energy splittings of the time-modulated system agree with those of the integrable approximation, showing that the dominant transport mechanism from one island to the other is direct tunnelling.

Starting from the QRR and decreasing the effective Planck's constant, we reach a regime where $\kbar$ is comparable to the phase space area of the period-one islands of regular motion. If chaos is present in the classical phase space, the dominant tunnelling mechanism is then expected to be chaos-assisted tunnelling as described earlier. According to Podolskiy and Narimanov \cite{Podolskiy2003}, the tunnel splittings yielded by such a mechanism have the following dependence on $\kbar$:
\begin{equation}\label{eq-PN}
\Delta{E}\propto\kbar\frac{\Gamma\left({A}/{\pi\kbar},{2A}/{\pi\kbar}\right)}{\Gamma\left({A}/{\pi\kbar}+1,0\right)}\textrm{,}
\end{equation}
where $A$ is the phase-space area of one island of regular motion and $\Gamma$ the {upper} incomplete Gamma function \cite{Abramowitz}. We observe the transition from the QRR to this regime in Fig.~\ref{fig-transition}, where the tunnelling behaviour of $H_m$ diverges from the {approximate exponential behaviour} yielded by $H_i$ to follow the Podolskiy-Narimanov theory as $\kbar$ is decreased. A more generic result than Eq.~(\ref{eq-PN}) for the chaos-assisted tunnelling splitting has recently been found by B{\"a}cker {\it et al.} \cite{Baecker2008}.

However, neither the Podolskiy-Narimanov model nor the B{\"a}cker {\it et al.}~model takes into account fine features of the classical phase space, for example partial dynamical barriers embedded in the sea of chaos \cite{Geisel1986}, or the internal structure of the islands of regular motion themselves. Therefore, even if some systems are well described by these models for a certain range of $\protect\kbar$, decreasing this parameter and therefore the resolution of the quantum coarse-graining can make the system ``see'' those small structures. As can be seen in Fig.~\ref{fig-kbargraph}, this leads to dramatic deviations from Eq.~(\ref{eq-PN}).

For $\protect\kbar \alt 0.1$, it is once again possible to explain the observed deviations by a change in the dominant transport mechanism from one island to the other. It is known that in the semiclassical limit $\kbar\rightarrow 0$, tunnelling through regular phase-space is greatly facilitated by the presence of resonance chains \cite{Brodier2001,Brodier2002}. It is therefore expected that for low $\kbar$ tunnelling between an island of regular motion and the sea of chaos will be enhanced by the presence of a secondary resonance chain embedded in this island.

A model reflecting this idea has been developed by Eltschka and Schlagheck \cite{Eltschka2005}, with a detailed description given in Ref.~\cite{Mouchet2006}. Let us assume each of the symmetric islands of regular motion supports a $r$:$s$ secondary resonance chain --- i.e. where $s$ internal oscillation periods match $r$ driving periods, and $r$ sub-islands are visible on the Poincar\'e section. This situation is well illustrated {in our system for the parameters} presented in Fig.~\ref{fig-kbargraph} (see inset). Firstly, the authors approximate the dynamics within the island {to that of} a pendulum and treat the pendulum potential $2V_{r:s}\cos{r\theta}$ as a perturbation. This perturbation couples the ground state $|\psi_0\rangle$ of the unperturbed approximate Hamiltonian only to the excited states $|\psi_{lr}\rangle$ where $l$ is an integer. In the following, we will denote $|\phi_{l}\rangle=|\psi_{lr}\rangle$. Denoting by $2\pi I_c$ the phase-space area of the island of regular motion (measured on a Poincar\'e section), they calculate the coupling of the unperturbed ground state of the island to its $k$th $\phi$-type state through this resonance, where $k$ is defined by $I_{(k-1)}<I_c<I_{k}$ ($I_l$, $l\in\mathbb{N}$ denotes the action associated with the $l$th $\phi$-state). In other words, they calculate the coupling of the {approximate ground state of the island} to the lowest $\phi$-state localised in a region where chaos is present for the exact Hamiltonian. They then assume that this coupling describes the effective coupling of $\phi_0$ to the sea of chaos and model the latter by two matrices of the Gaussian orthogonal ensemble (one for each parity), which means that they neglect the effects of partial barriers in the chaotic part of phase space. Using the results from Ref.~\cite{Leyvraz1996}, they are therefore able to calculate an expectation value for the logarithm of the quasi-energy splittings for each set of parameters, which we compare to our numerical results in Fig.~\ref{fig-kbargraph}.

We find reasonable qualitative and quantitative agreement in the range $\kbar<0.1$, which is expected as this is precisely the order of magnitude of the phase-space area of the dominant chain's resonances. We also observe an effect already noted by Eltschka and Schlagheck \cite{Eltschka2005}, namely the fact that the sharp decrease of the analytically calculated tunnelling splittings is shifted to lower values of $\kbar$ as compared to the similar decrease of the splittings obtained from the numerical simulations. Such steps occur when the value of the index $k$ changes, i.e. every time an unperturbed $\phi$-state of the approximate Hamiltonian crosses the somewhat artificial limit $I_c$ of the island of regular motion. This happens for every integer $l$ such that $I_c=\kbar(lr+1/2)$ (in our case, $r=6$ and $l=1$). It is known, however, that states that are neither chaotic nor regular (\emph{hierarchical} states: see Ref.~\cite{Ketzmerick2000}) can be localised in the vicinity of the island. Therefore, the ``effective value'' of the islands' phase-space area could be a little higher than $2\pi\times I_c$. This could account for the observed shift of the step.

{Recently a unified description that can predict tunnel splittings from the direct tunnelling regime to the resonance assisted regime has been developed \cite{Loeck2010}.  It is hopeful that this theory would provide an improved description of the the data in Fig.~\ref{fig-kbargraph}, however its complexity puts it beyond the scope of this paper.}

\section{Prospects for experiments}
\label{sec:exp}

We now examine the feasibility of performing dynamical tunnelling experiments that explore the effects presented in the previous sections with a Bose-Einstein condensate in a magnetic microtrap. To this end we discuss the realisation of a one-dimensional Hamiltonian in \sref{sec-confinement}, realistic parameters for the trapping potential in \sref{sec:prospects}, the effects of atomic interactions in \sref{sec:meanfield}, the initial state creation and loading in \sref{sec:inistate}, and finally condensate sizes and inelastic losses in \sref{sec:othernumbers}. 

\subsection{Confinement geometry}
\label{sec-confinement}

\begin{figure}
\begin{center}
\includegraphics[width=8.6cm]{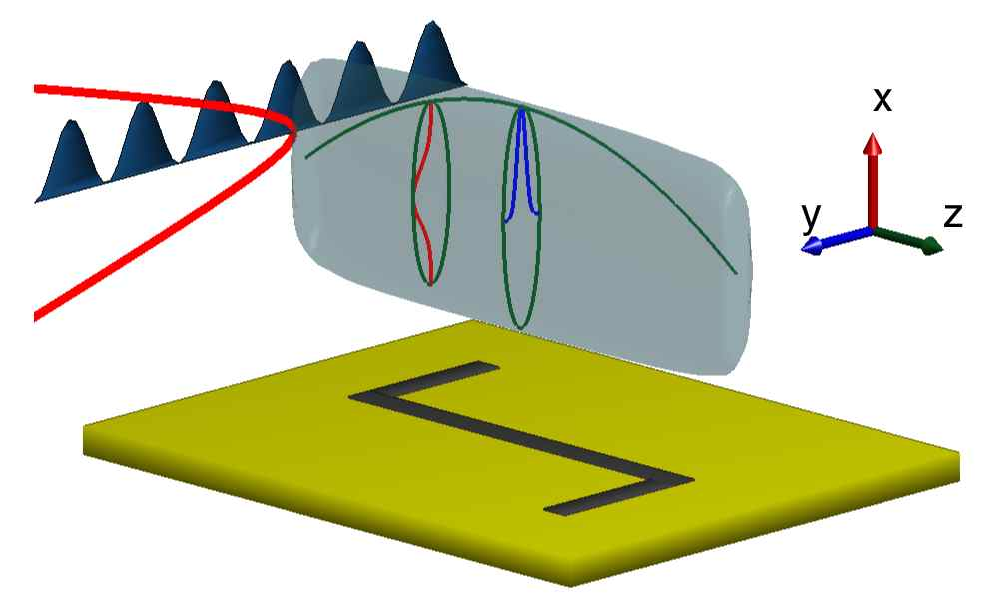}
\caption{(Color online) Schematic of the arrangement of condensate, confining potentials and atom-chip surface (yellow) with characteristic z-shaped wire. The atom chip potential has the form \eref{eqn:H1text} along $x=q$ (red) and is weakly harmonic along $z$ (potential not shown). Superimposed is an optical lattice potential (blue), freezing out the dynamics along $y$. The gray shaded volume is the resulting isodensity surface of a trapped BEC. Embedded density profiles along the $x$, $y$, $z$, co-ordinate directions are also sketched.}
\label{fig-expsetup}
\end{center}
\end{figure}

A key difficulty is that the form of the Hamiltonian Eq.~(\ref{eqn:H1text}) is  realized in terms of the two-dimensional (2D) radial coordinate $r=\sqrt{x^2 + y^2}$ of the atom-chip trap, see \aref{sec:trap}. Instead, we require it to describe a 1D degree of freedom, say $q=x$. Without this constraint, the islands of classically bounded regular motion can be dynamically linked via the second radial dimension ($y$). Classical diffusion can then populate the other island and mask the signature of dynamical tunnelling. While the tunnelling and diffusion rates will in general be quite different, it would be desirable to avoid this effect altogether by freezing out the dynamics in the $y$-direction. This could be achieved by applying an optical lattice formed by two laser beams propagating along the $\pm y$ directions and confining the atoms in one well of the lattice, as sketched in \fref{fig-expsetup}.

In contrast to the $y$-dimension, the $z$-dimension can be very weakly confined on an atom-chip. The condensate in that direction then has  an approximately ``infinite'' extension. We will check this assumption below in \sref{sec:othernumbers}.

\subsection{Accessible tunnelling parameters}
\label{sec:prospects}

We begin by proposing a regime in which it is feasible to observe dynamical tunnelling. From a practical point of view, we wish to minimise the tunnelling period and the sensitivity of the tunnelling to any small fluctuations in experimental parameters. Distinguishable period-one islands appear in the Poincar\'e section for $\kappa > 1$, and the closer the islands are in phase space, the shorter the tunnelling period.  However, if the two islands are too close to each other, separating them in time-of-flight absorption images could be challenging.  High values of $\kbar$  yield faster tunnelling, but require very tight trapping potentials.   A convenient and accessible choice of parameters within the QRR is $\kappa=1.01$, $\kbar=1/73$, and $\epsilon = 0.48$, which gives a relatively short tunnelling period of $159$ modulations of the trapping potential.  This has the additional advantage that the tunnelling period is relatively insensitive to small, uncontrolled experimental variations in the exact value of the parameters, with the strongest dependence being on $\kappa$. We calculate that the following variations would induce shifts of less than $10\%$ in the tunnelling period:
\begin{equation}
1.009<\kappa<1.011,\quad 1/77<\kbar<1/69,\quad 0.37<\epsilon<0.59.
\end{equation}

\begin{figure}
\begin{center}
\includegraphics[width=7.6cm]{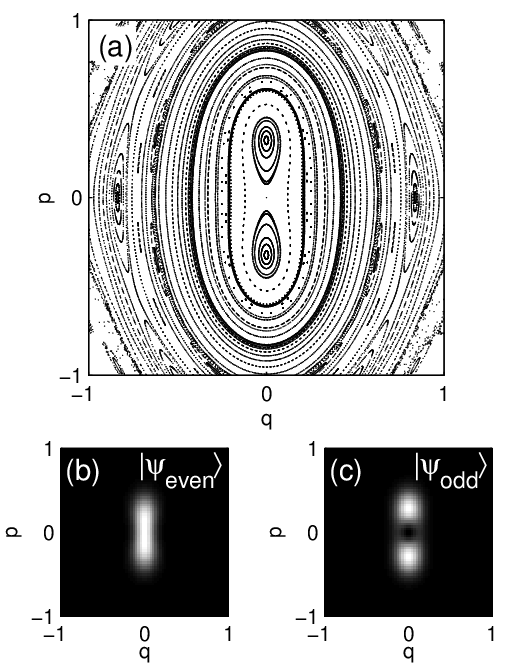}
\caption{(a) Poincar\'e section at $t=0$ for the proposed experimental parameters $\kappa=1.01$, $\kbar=1/73$,  $\epsilon = 0.48$. Note that no chaos is present between the islands. (b) Husimi function for the even tunnelling state.  (c) Husimi function for the odd tunnelling state.}
\label{fig-expPoinc}
\end{center}
\end{figure}

The Poincar\'e section and Husimi functions of the two tunnelling Floquet states are shown in Fig.~\ref{fig-expPoinc}. 
To achieve this with a BEC of $^{87}$Rb atoms in the $F=2$, $m_F=2$ magnetic sub-state confined on an atom chip would require a current through the trapping wire of $I=63~$mA and a bias magnetic field of $B_b = 5~$G. In the limit of an infinitesimal wire, this yields a trap $25~\mu$m from the wire
which, when combined with an axial magnetic field of $B_0 = 0.25$ G, has a radial trapping frequency in the harmonic approximation of $\omega_r = 2\pi \times 5~$kHz.  At such {distances} wire imperfections that could distort the confining potential and cause cloud fragmentation can be avoided~\cite{Kruger2007}. The trap depth is of the same order of magnitude as the atomic Zeeman energy from the magnetic field due to the wire current at the location of the trap minimum. For our parameters it is roughly equal to $\sub{E}{mag} =  g m_F \mu_B B_b\approx 0.3$ mK, more than sufficient to trap a BEC at nanokelvin temperatures. 

\subsection{Bose-Einstein condensates and mean-field effects}
\label{sec:meanfield}

So far we have considered the physics of single atoms within the driven potential. This has previously been sufficient to describe experiments in very dilute thermal or condensed gases \cite{Hensinger2001,Steck2001}. 
However, to work with a trapping potential with a single minimum, to controllably load small regions of phase space, and to still be able to image the resulting dynamics, it is necessary to use atoms that are sufficiently cold and dense that they are Bose condensed.
In realistic BECs  nonlinear mean-field interactions arising from atomic $s$-wave collisions can play an important role in the dynamics~\cite{stringari:review}. These can be described using the
one-dimensional Gross-Pitaevskii equation (GPE)
\begin{align} i\kbar \frac{\partial}{\partial t} \psi& =\left[ -\frac{\kbar^{2}}{2}\frac{\partial^{2}}{\partial q^{2}} + V(q,t) + U_{\textrm{1D}} |\psi|^{2} \right]\psi.
\label{standardgpe}
\end{align}
Here the condensate wavefunction $\psi(q,t)$ is  normalized to one and the potential is $V(q,t)=\kappa(1+\epsilon \cos(t))(1+q^{2})^{1/2}$. The relation between the effective one-dimensional interaction-strength $U_{\textrm{1D}}$ and the physical parameters of the system is derived in \aref{sec:gpe}. 

\begin{figure}
\begin{center}
\includegraphics[width=8.6cm]{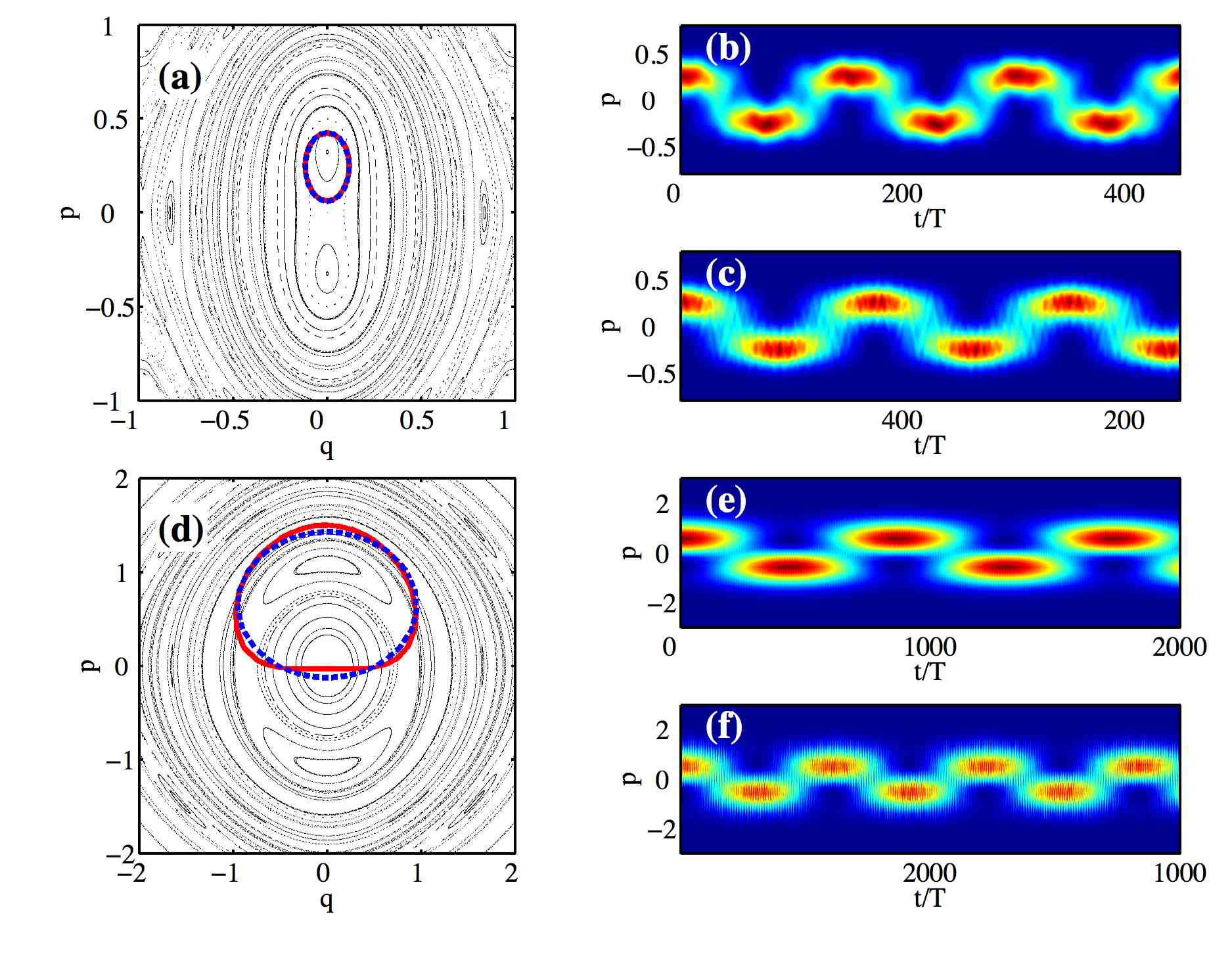}
\caption{(Color online) (a--c) Dynamical tunnelling signatures from condensate mean-field theory according to \eref{standardgpe}, for parameters as in \fref{fig-expPoinc} and $U_{\textrm{1D}}=1\times 10^{-4}$. (a) Husimi function (FWHM) of the initial state located within the Poincar\'e section: (thick red, solid) initial Floquet state $|\phi_+ \rangle$. (thick blue, dashed) experimentally accessible Gaussian approximation as explained in the text, $|\varphi_+ \rangle$. (b) Dynamical tunnelling in momentum space from initial Floquet state and (c) from initial Gaussian approximation as explained in the text. (d--f) The same as (a--c) but for parameters $\kappa=1.3$, $\kbar=0.5$, $\epsilon = 0.2$ and $U_{\textrm{1D}}=0.01$, with $p_0=0.65$ and $\sub{\kappa}{ini}=0.4$.  }
\label{fig-gpetunnelling}
\end{center}
\end{figure}
We have performed numerical simulations of the 1D Gross-Pitaevskii equation~(\ref{standardgpe}) with the BEC intially loaded  in the pure Floquet superposition state $|\phi_+\rangle$. 
For the case described in \sref{sec:prospects}, we find that dynamical tunnelling oscillations can be observed with \eref{standardgpe} for nonlinearities up to about $U_{\textrm{max}}=1.8\times 10^{-4}$, beyond which the nonlinearity shuts down dynamical tunnelling.  For these particular parameters we have found that this is an example of macroscopic quantum self-trapping, which we have analysed in detail elsewhere \cite{wuester:nonlinearDT}.  For the purposes of this paper, we choose $U_{\textrm{1D}}=1\times 10^{-4}$ which results in unhindered dynamical tunnelling oscillations as demonstrated in \fref{fig-gpetunnelling}(b), where we show the momentum-space distribution at integer multiples of the modulation period. This momentum-space image constitutes a direct experimental observable, as standard time-of-flight expansion and absorption imaging techniques convert the centre of mass momentum into  spatial positions that are easily resolved  \cite{book:pethik}. 
An improved signal to noise can potentially be obtained by imaging along the weakly-trapped $z$-direction, as the atomic density will be spread across fewer pixels on the camera. 

Finally, we note that the onset of suppression of dynamical tunnelling by nonlinearities depends strongly on the system parameters, and in some regimes does not occur at all.  We have studied the presence and absence of macroscopic quantum self-trapping in Ref.~\cite{wuester:nonlinearDT}.

\subsection{Experimental and numerical initial states}
\label{sec:inistate}

In this section we consider how to best load the initial Floquet superposition  $|\phi_+\rangle$ experimentally, as well as the effects of imperfect loading.
In \fref{fig-gpetunnelling}(c) we show the results of a simulation with the same parameters as for \fref{fig-gpetunnelling}(b), but with the initial state being a BEC dynamically prepared from the ground state of the trapping potential such that it obtains a large overlap with the  Floquet superposition  $|\phi_+ \rangle$. We label the resulting state $|\varphi_{+} \rangle$.  Numerically, we create $|\varphi_{+} \rangle$ as follows:
(i) We assume an initial potential strength $\sub{\kappa}{ini}$, different from $\kappa$. In the potential $V(q) = \sub{\kappa}{ini}(1+\epsilon)(1+q^{2})^{1/2}$, we determine the condensate ground state using imaginary time evolution \cite{book:pethik}. (ii) The resulting wave function $f_1(q)$ is given a momentum kick $f_2 = f_1\exp{[i p_0 \kbar t]}$. (iii) We vary $\sub{\kappa}{ini}$ and $p_0$ to maximize the overlap $\langle\phi_+|f_2\rangle$ and finally take this as the initial state $|\varphi_+ \rangle[ \sub{\kappa}{ini},p_0]$ for our simulation. We see the dynamical tunnelling  arising from this procedure in \fref{fig-gpetunnelling}(c), where  we have used $p_0=0.244$ and $\sub{\kappa}{ini}=4.5$.

The corresponding experimental sequence is as follows: we  begin with a stationary BEC trapped in the ground state of the magnetic potential with trap strength given by $\sub{\kappa}{ini}$. The wire current would then be slowly decreased by an amount $\delta I = 0.78~$mA and then suddenly switched back to $63$ mA to start the cloud oscillating about the centre of the trap minimum, acquiring the maximal momentum $p_0$. When the BEC passes the bottom of the trap for the first time ($t = 0$), the simultaneous modulation of wire current and bias magnetic fields at $\epsilon=0.48$ and $\Omega=2\pi\times 4.975\,\textrm{kHz}$ begins, resulting in the loading of an equal superposition of the odd and even tunnelling Floquet states with a fidelity of up to 95\%. At this moment the trap-strength is also changed to $\kappa$. Following the described sequence, the state $|\varphi_+ \rangle$ can be directly created in the experiment.

On closer inspection, the simulations beginning with this initial state show a slightly different dynamical tunnelling period compared to those that begin from a Floquet state superposition. We find such modifications in the presence of nonlinear interactions whenever the initial state slightly differs from the exact Floquet state $|\phi_+ \rangle$. As we vary $\sub{\kappa}{ini}$ and $p_0$ used to define $|\varphi_+ \rangle$, the tunnelling period undergoes small but continuous changes. This effect is absent in the linear case with $U=0$. 

\subsection{Condensate parameters}
\label{sec:othernumbers}

The parameters of \fref{fig-gpetunnelling}(a--c) amount to a condensate of $N=171$ atoms [\eref{NofU}], assuming the axial extension of the condensate is $50$ $\mu$m, with a corresponding peak density of $2\times 10^{14}$ $\textrm{cm}^{-3}$.  
For comparison, we show in \fref{fig-gpetunnelling}(d--f) a second parameter set with much larger $\kbar = 1/2$, with other parameters $\kappa=1.3$, $\epsilon = 0.2$ and $U_{\textrm{1D}}=0.01$, corresponding to $N=20$ atoms. Realising $\kbar=1/2$ would either require us to lower $B_0$ to $17$ mG or increase $\omega_x$ to  about $2\pi\times 300$ kHz.  While challenging, precise imaging of small clouds of atoms as discussed here has been achieved \cite{Esteve2008a,Maussang2010a}, however the reliability of mean-field theory in this regime is questionable.

We estimate that three-body losses in this scenario would limit the BEC lifetime to a few hundred milliseconds \cite{Soding1999} which is sufficiently long to conduct the experiment.  The Majorana transition-induced losses will be negligible as $\omega_{\textrm{Larmor}}/\frac{d\theta}{dt} \gtrsim 10$ at all times, where $\theta$ is the angle between the magnetic field and the $z$ axis at the position of a classical period-one resonance.

As mentioned in \sref{sec-confinement}, the freezing of the $y$-direction could be achieved by applying an optical lattice formed by two laser beams propagating along the $\pm y$ directions and confining the atoms in one well of the lattice. The transverse (radial) size of the condensate with $171$ atoms is less than $200$ nm, which would easily fit into a single lattice site. Therefore, not only would dynamical tunnelling take place exclusively in the plane perpendicular to the lattice, it would also occur in a single lattice site.

Finally, we have assessed the influence of the trapping potential in the $z$-dimension, and of experimental imperfections in the trap alignment. We have explicitly verified that the weakly confined $z$-dimension can be ignored, as mentioned in \sref{sec-confinement}. To this end, we modelled the dynamics of \fref{fig-gpetunnelling} using the full two-dimensional GPE, \eref{scaled2Dgpe}, with a  weak harmonic trap in the $z$-direction. The axial condensate wave function had a non-trivial Thomas-Fermi profile \cite{book:pethik}. The tunnelling signal in momentum-space is as clear as those shown in \fref{fig-gpetunnelling}(b,c), even accounting for imaging along the $z$-axis.  Furthermore, in two dimensional simulations we have found the dynamical tunnelling to be robust to changes in the initial state away from the ideal Floquet superposition, as well as small offsets from the trap centre.

Unfortunately, it seems that the tight trapping potentials combined with the small atom numbers present a significant experimental challenge for dynamical tunnelling with BECs to be realised as described here.  The  small atom numbers found for these examples are mostly due to the need to freeze out dynamics in the $y$-direction.
An investigation to what extent dynamical tunnelling can still be identified if dynamics in the $y$-direction is retained would be of interest, but is beyond the scope of the present paper. Note that the small number of atoms is required in order to keep the peak-density low enough for the effect of interactions to remain small. The effect of stronger interactions can be quite complicated, but elsewhere we have found that it is possible to observe dynamical tunnelling in regimes with much larger atom numbers  \cite{wuester:nonlinearDT}.

\section{Conclusions}\label{sec:concl}
We have studied dynamical tunnelling in a driven 1D single-well potential {provided by the magnetic field of}  a current-carrying wire on an atom chip. An experiment performed in this geometry would provide the possibility of studying the quantum-classical transition as a function of the effective Planck's constant.  It also has a clear correspondence to the classical dynamical picture,  uncomplicated by coherences between neighbouring wells of a standing wave~\cite{Hensinger2001}.

We have analyzed the dependence of the dynamical tunnelling rate with the experimental parameters and observed that it fluctuates considerably, a feature previously observed for other trapping potentials. In the limiting case of the quantum regular regime, we were able to relate the variation in the tunnelling rate to the energy spectrum of an integrable Hamiltonian. In the classical chaotic regime, the variation in the tunnelling rate is linked to a chaos-assisted tunnelling phenomenon for high $\kbar$.   For lower $\kbar$ we have underlined the role of classical resonances linking the islands of regular motion with the sea of chaos.

Finally we have considered the experimental prospects for realising dynamical tunnelling with a BEC in a magnetic microtrap. We have found that dynamical tunnelling would be observable on a $10$ ms timescale for realistic atom-chip parameters provided the dynamics in the plane perpendicular to the motion could be frozen out.  However, this requires BECs with a rather small number of atoms. Our conclusions  hold in the presence of nonlinearities due to atomic interactions as long as these are not too large.

\acknowledgments{ This research was supported under the Australian Research Council's Discovery Projects funding scheme (project numbers DP0343094 and DP0985142) and by the ARC Centre of Excellence for Quantum-Atom Optics (project number CE0348178).  G.~J.~M. acknowledges the support of the Australian Research Council grant FF0776191.  M.~L. would like to thank Dominique Delande, Christopher Eltschka and Peter Schlagheck for stimulating discussions. We would like to thank Tod Wright for a careful reading of the  manuscript.}

\appendix

\section{Atom-chip Hamiltonians}
\label{sec:trap}

We consider atoms trapped in a magnetic potential
formed by a current flowing  in the {positive $z$
direction} along single infinite conductor, with a constant bias field $B_{b}$ in the {negative} $y$ direction. This forms
a 2D potential in the $x$ and $y$ dimensions. For now we will ignore any dynamics
in the $z$ dimension --- this is reasonable as on typical atom chips the trapping potential in this dimension is much weaker and the dynamics {are correspondingly} slower.


The magnetic field strength a distance $\sqrt{x^2 + y^2}$ from the conductor will be
\begin{equation}
|B(r)| = \frac{\mu_0}{2\pi} \frac{I}{\sqrt{x^2 + y^2}}.
\end{equation}
The bias field $B_b$ cancels the field from the wire along the line defined by
\begin{equation}
x = x_0 = \frac{\mu_0 I }{2 \pi B_b},
\end{equation}
where the field gradient will be
\begin{equation}
|B'| = \frac{\mu_0}{2\pi} \frac{I}{x_0^2} = \frac{B_b}{x_0}.
\end{equation}
We define $x'=x-x_0$ and $r = \sqrt{x'^2+y^2}$. If we add a small, possibly time-dependent offset field in the $z$ dimension of magnitude $B_0$ then the magnitude of the field at a distance $r$ from the minimum can be approximated by
\begin{equation}
B = \sqrt{ (r B')^2 + B_0^2},
\end{equation}
which will be valid for $r \ll x_0$ (for more detail see the next section). The radial potential that
the atoms see will be
\begin{equation}
V(r) = g m_F \mu_B B_0 \sqrt{ (B'/B_0)^2 r^2 + 1},
\end{equation}
where $g$ is the Land\'e $g$-factor, $m_F$ is the magnetic sublevel that is trapped and $\mu_B$ is the Bohr magneton.
For $(B'/B_0)^2 r^2 \ll 1$ a Taylor expansion gives
\begin{equation}
V(r) = \frac{1}{2} g m_F \mu_B \frac{B'^2}{B_0} r^2 + \mathrm{const.},
\end{equation}
i.e. the {minimum of the} potential is {approximately} harmonic with a radial trapping frequency of
\begin{equation}
\omega_r = B' \sqrt{\frac{g m_F \mu_B}{m B_0}}.
\end{equation}
The system Hamiltonian is then
\begin{equation}
\label{hamil:dimfull}
H = \frac{p^2}{2m} + g m_F \mu_B B_0 \sqrt{ (B'/B_0)^2 r^2 + 1}.
\end{equation}
{We define the} dimensionless quantities
\begin{equation}\label{eq-rescaling}
\tilde{r} = \frac{B'}{B_0} r, \qquad \tau = \Omega t,
\qquad\tilde{p} = \frac{B'}{B_0 m \Omega} p, \qquad
\tilde{H} = \frac{B'^2}{m \Omega^2 B_0^2} H
\end{equation}
{where $\Omega$ will be the angular frequency of the modulation.} {After} dropping tildes, we find our dimensionless Hamiltonian is
\begin{equation}
H = \frac{p^2}{2} + \kappa  \sqrt{  r^2 + 1},
\end{equation}
where we have defined the dimensionless parameter
\begin{equation}
\kappa = \frac{g m_F \mu_B B'^2}{m B_0 \Omega^2} \equiv
\frac{\omega_r^2}{\Omega^2} .
\end{equation}
The effective Planck's constant for this system is
\begin{equation}
\label{kbareqn}
\kbar = \frac{\hbar B'^2}{B_0^2 m \Omega} \equiv \sqrt{\kappa} \frac{\hbar \omega_r}
{g m_F \mu_B B_0}.
\end{equation}

\subsection{Modulation of the trapping potential}
We now consider the possible ways we can {modulate} the trapping potential {for} the atoms.  The magnetic field at location $(x,y)$ is given by
\begin{equation}
\mathbf{B} = \left\{-
\frac{yB_w(t)}{\sqrt{x^2 + y^2}}, \frac{x B_w(t)}{\sqrt{x^2 + y^2}} - B_b(t), B_0(t)\right\},
\end{equation}
where $B_0(t)$ is the offset field along the $z$-direction, $B_w(t)$ is the field from the 1D conductor, and $B_b(t)$ is the bias field along the y-direction.  We have
\begin{eqnarray}
B_b(t) &=& B_{b0} (1 + f_b), \\
B_0(t) &=& B_{0} (1 + f_z), \\
B_w(t) &=& \frac{\mu_0}{2\pi} \frac{I_0}{\sqrt{x^2 + y^2}} (1 + f_w),
\nonumber\\
&=& \frac{x_{0} B_{b0}}{\sqrt{x^2 + y^2}}(1 + f_w),
\end{eqnarray}
where we have used the shorthand $f_a = f_a(t)$ {for any modulation for $t>0$}. We also
define the constant
\begin{equation}
x_0 =  \frac{\mu_0 I_0 }{2 \pi B_{b0}},
\end{equation}
which is the location of the trap minimum with no driving.

The magnitude of the magnetic field can be shown to be
\begin{align}
B^2 &= B_0^2(1 + f_{z})^2  + \frac{B_{b0}^2}{(x_{0} + x')^2 + y^2}\big[
(1+f_{b})^2(x'^{2} +y^2)
\nonumber\\
 &+ (f_{b} -f_{w})^2 x_{0}^2  + 2 (1+f_{b})(f_{b} - f_{w})x'x_{0}
\big].
\end{align}
where $x'=x-x_0$. 

\subsection{Choice of time-dependent modulations}
Various Hamiltonians can be realised with appropriate choices of modulation of the magnetic fields and currents.  Here we detail two that are potentially of interest.
\begin{enumerate}
\item $f_b = f_w = f_z$:\\
In this case the trap minimum is stationary, and both the field gradient and the offset field $B_0$ are modulated. This yields the magnetic field
\begin{equation}
B = B_0 (1+ f_b)\left(1 +  \tilde{r}^2\right)^{1/2},
\end{equation}
which has a similar form to the modulated standing wave used in Refs.~\cite{Steck2001,Hensinger2001} --- a potential with a stationary minimum but a modulated strength.

\item $f_b = 0$,  $f_z=0$:\\
This case is probably the easiest experimentally, as only the current in the
trapping wire needs to be modulated.  This avoids any difficulties with
inductances in the coils providing the bias and offset magnetic fields.  The field in this case is
\begin{equation}
B = B_0 \left\{1 + \left(\frac{B'}{B_{0}} \right)^2\left[y^2 + \left(x' - x_{0}f_w \right)^2  \right\}
\right]^{1/2}.
\end{equation}
This potential has a constant strength but the position of its minimum oscillates.
\end{enumerate}

In summary,  the experimentally relevant Hamiltonians in dimensionless units are
\begin{eqnarray}
H &=& \frac{p^2}{2} + \kappa (1+ \epsilon \cos\tau)\left[1 +  ( x'^2 + y^2)
\right]^{1/2},
\label{eqn:H1}
\\
H &=& \frac{p^2}{2} + \kappa \left[1 + y^2 + \left(x' - x_{0} \epsilon \cos\tau\right)^2
\right]^{1/2}.
\label{eqn:H2}
\end{eqnarray}
In this paper we have only considered Hamiltonian (\ref{eqn:H1}) as this is analogous to the experiments performed in the modulated standing wave.

\section{Dimensional reduction of the Gross-Pitaevskii equation}
\label{sec:gpe}

We begin with the standard Gross-Pitaevskii equation for the mean-field of a condensate tightly confined along the $y$-dimension \cite{stringari:review}
\begin{align}
i\hbar\frac{\partial}{\partial t}\psi(x,z)&=\Big[-\frac{\hbar^{2}}{2m}\left(\frac{\partial^{2}}{\partial x^{2}} + \frac{\partial^{2}}{\partial z^{2}}  \right)+ V(x,z,t) 
\CR
&+ \gamma_{\textrm{2D}} |\psi(x,z)|^{2} \Big]\psi(x,z),
\label{orig2Dgpe}
\end{align}
with
\begin{align}
V(x,z,t)&=V_{0} \big[ 1 + \epsilon \cos(\Omega t) \big]  \sqrt{1+ (x/d)^{2}  },
\label{chip_potential}
\end{align}
The effective 2D interaction is $\gamma_{\textrm{2D}}=2\sqrt{2\pi}\hbar^{2} a_{s}/(m a_{y})$ \cite{Subasi2009}, where $a_{s}$ is the 3D scattering length and $a_{y}=\sqrt{\hbar/(m\omega_{y})}$ the oscillator length in the frozen $y$-direction. $\int dxdz |\psi|^2=N$, the total 3D number of atoms. The parameters of the potential can be read from \eref{hamil:dimfull} with the identification $x\leftrightarrow r$.

We now measure length in units of $d$, time in units of $\Omega^{-1}$ [see \eref{dimlessunitssscales}] and employ the dimensionless wave function $\tilde{\psi}= \psi \sqrt{d/N}$, which is normalized to one. Finally relabelling $\tilde{\psi}\rightarrow \psi$ we obtain:
\begin{align}
i \kbar\frac{\partial}{\partial t}\psi&= \Big[-\frac{\kbar^{2}}{2}\left(\frac{\partial^{2}}{\partial x^{2}} + \frac{\partial^{2}}{\partial z^{2}}  \right) + V(x,z,t) 
\CR
&+ U_{\textrm{2D}} |\psi(x,z)|^{2} \Big]\psi,
\label{scaled2Dgpe}
\end{align}
where 
\begin{align}
V(x,z,t) = \kappa [1+ \epsilon \cos(t)]\sqrt{1 + x^2} .
\label{potential}
\end{align}
Importantly $U_{\textrm{2D}}=N \gamma_{\textrm{2D}} /m \Omega^{2}d^{4}$. We now eliminate the dimension $z$, assuming the size of the condensate in that direction, $L_{z}$, to be large and the 
dynamics hence slow. We simply set $\psi(x,z) =  \tilde{\psi}(x) \phi(z)$, with $\phi(z)=\theta(\tilde{L}/2 - |z|)/\sqrt{\tilde{L}}$, where $\theta$ is the Heaviside function and $\tilde{L}=L_{z}/d$.
Insertion into \eref{scaled2Dgpe} yields:
\begin{align}
i \kbar\frac{\partial}{\partial t}\tilde{\psi}&=\left[-\frac{\kbar^{2}}{2}\frac{\partial^{2}}{\partial x^{2}}  + V(x,t) + U_{\textrm{1D}} |\tilde{\psi}(x)|^{2} \right]\tilde{\psi},
\label{scaled1DGPE}
\end{align}
with $U_{\textrm{1D}}=U_{\textrm{2D}}/L_{z}$. We obtain \eref{standardgpe} by replacing $x\rightarrow q$ and have
\begin{align}
U_{\textrm{1D}}&=\frac{N\gamma_{\textrm{2D}}}{m\Omega^{2} d^{3} L_{z}} = \frac{2\sqrt{2\pi} \hbar^{2}a_{s}N }{m^{2} a_{y} \Omega^{2} d^3 L_{z}} =\frac{2\sqrt{2\pi} \hbar^{2}a_{s}N \kappa }{m a_{y} d L_{z}V_{0}}.
\label{UofN}
\end{align}
To rewrite \eref{UofN} in terms of the quantities $\kbar$, $\kappa$, $\epsilon$ and $U_{\textrm{1D}}$ that underly most of our numerical results, 
we substitute $d=B_0/B'$, $V_0=g m_F \mu_B B_0$ [comparing \eref{chip_potential} and \eref{hamil:dimfull}] and then use \eref{kbareqn}
to arrive at:
\begin{align}
N&=\frac{L_z U_{\textrm{1D}}}{2\sqrt{2\pi} a_s\sub{\hbar}{eff}^{3/2}\kappa^{1/4}}\sqrt{\frac{\omega_x}{\omega_y}}.
\label{NofU}
\end{align}
The atom numbers and densities quoted in \sref{sec:meanfield} correspond to $L_{z}=50$ $\mu$m, $\omega_{z}=2\pi\times18.6$ Hz, $\omega_{y}=2\pi\times30$ kHz and other parameters as given in \sref{sec:prospects}.


\end{document}